\documentclass[amsbook,12pt]{article}
\usepackage{epsfig, amsfonts,amssymb, amsmath}
\usepackage[usenames]{color}
\usepackage{picinpar}
\usepackage{float, graphicx}
\usepackage{hyperref}
\usepackage{hyperref}
\newcommand{\nl}{\mbox{}\\}

\begin{document}
\thispagestyle{empty}
%
%
\mbox{} \vspace{-0.750cm} \\
\hypersetup{
colorlinks=true,
linkcolor=blue,
filecolor=magenta,
urlcolor=blue,
}
\urlstyle{same}
%
%
\begin{center}
%
%
{\Large \bf A matlab code to compute reproduction numbers} \\
\mbox{} \vspace{-0.250cm} \\
{\Large \bf with applications to the Covid-19 outbreak} \\
\nl
\mbox{} \vspace{-0.300cm} \\
{\sc Paulo R. Zingano, Jana\'\i na P\!\:\!. Zingano,} \\
\mbox{} \vspace{-0.575cm} \\
{\footnotesize Institute of Mathematics and Statistics} \\
\mbox{} \vspace{-0.700cm} \\
{\footnotesize Universidade Federal do Rio Grande do Sul} \\
\mbox{} \vspace{-0.700cm} \\
{\footnotesize Porto Alegre, RS 91509-900, Brazil} \\
\mbox{} \vspace{-0.150cm} \\
{\sc Alessandra M. Silva} \\
\mbox{} \vspace{-0.575cm} \\
{\footnotesize Companhia de Planejamento do Distrito Federal} \\
\mbox{} \vspace{-0.700cm} \\
{\footnotesize Governo de Bras\'\i lia} \\
\mbox{} \vspace{-0.700cm} \\
{\footnotesize Bras\'\i lia, DF 70620-080, Brazil} \\
\mbox{} \vspace{-0.350cm} \\
{\sc and} \\
\mbox{} \vspace{-0.350cm} \\
{\sc Carolina P\!\:\!. Zingano,} \\
\mbox{} \vspace{-0.600cm} \\
{\footnotesize School of Medicine} \\
\mbox{} \vspace{-0.700cm} \\
{\footnotesize Universidade Federal do Rio Grande do Sul} \\
\mbox{} \vspace{-0.700cm} \\
{\footnotesize Porto Alegre, RS 90035-003, Brazil} \\
\nl
\mbox{} \vspace{-0.250cm} \\
{\bf Abstract} \\
\begin{minipage}[t]{12.750cm}
{\small
\mbox{} \hspace{+0.250cm}
We discuss the generation of various
{\em reproduction ratios\/} or {\em numbers}
that are very useful
to monitor an ongoing epidemic
like Covid-19
and examine the effects
of intervention measures.
A detailed SEIR algorithm
is described for their computation,
with applications given to
the current Covid-19
outbreaks in a number of countries
(Argentina, Brazil, France, Italy, Mexico,
Spain, UK and USA).
The corresponding {\em matlab script},
complete and ready to use,
is provided for free downloading.
}
\end{minipage}
\end{center}
\nl
\mbox{} \vspace{-0.250cm} \\
\mbox{} \hspace{+0.250cm}
\begin{minipage}[t]{14.000cm}
{\bf Key words:}
{\small
Covid-19 outbreak, SARS-Cov-2 coronavirus,
reproduction numbers, \\
\mbox{} \hfill
SEIR deterministic models,
parameter uncertainties,
robust methods \\
}
\end{minipage}
\nl
\mbox{} \vspace{-0.350cm} \\
\nl
\mbox{} \hspace{+0.250cm}
\begin{minipage}[t]{14.000cm}
{\bf Matlab code:}
{\small
%
%
\end{minipage}
\newpage
%
%
\mbox{} \vspace{-1.250cm} \\
\setcounter{page}{1}
%
%
%
%

{\bf 1. Introduction} \\

The monitoring of the evolving state of a serious
epidemic can be done during and after its outbreak
by estimating the daily values of basic ratios
generally known as reproductive or reproduction numbers
\cite{Driessche2008, Heffernan2005, Hethcote2000, Mellan2020}.
\!While not properly geared to allow
serious predictions of future values
of the epidemic,
they are nevertheless able to
display the past and present
history with amazing clarity.
\!However, as their calculation
depends on the values of various
mathematical parameters
(like the length of transmission
and incubation periods),
this ability may be impaired
by inaccuracies in
their estimation.
This is particularly true
for the
widely used
{\em basic reproduction number},
which measures the average number
of secondary cases
generated by a typical infectious individual
in a full susceptible population
(Figure~1). \\
\nl
%
%
%
%
%
\mbox{} \vspace{-0.450cm} \\
\mbox{} \hspace{-0.250cm}  
\begin{minipage}[t]{10.000cm}  
\includegraphics[keepaspectratio=true, width=9cm]{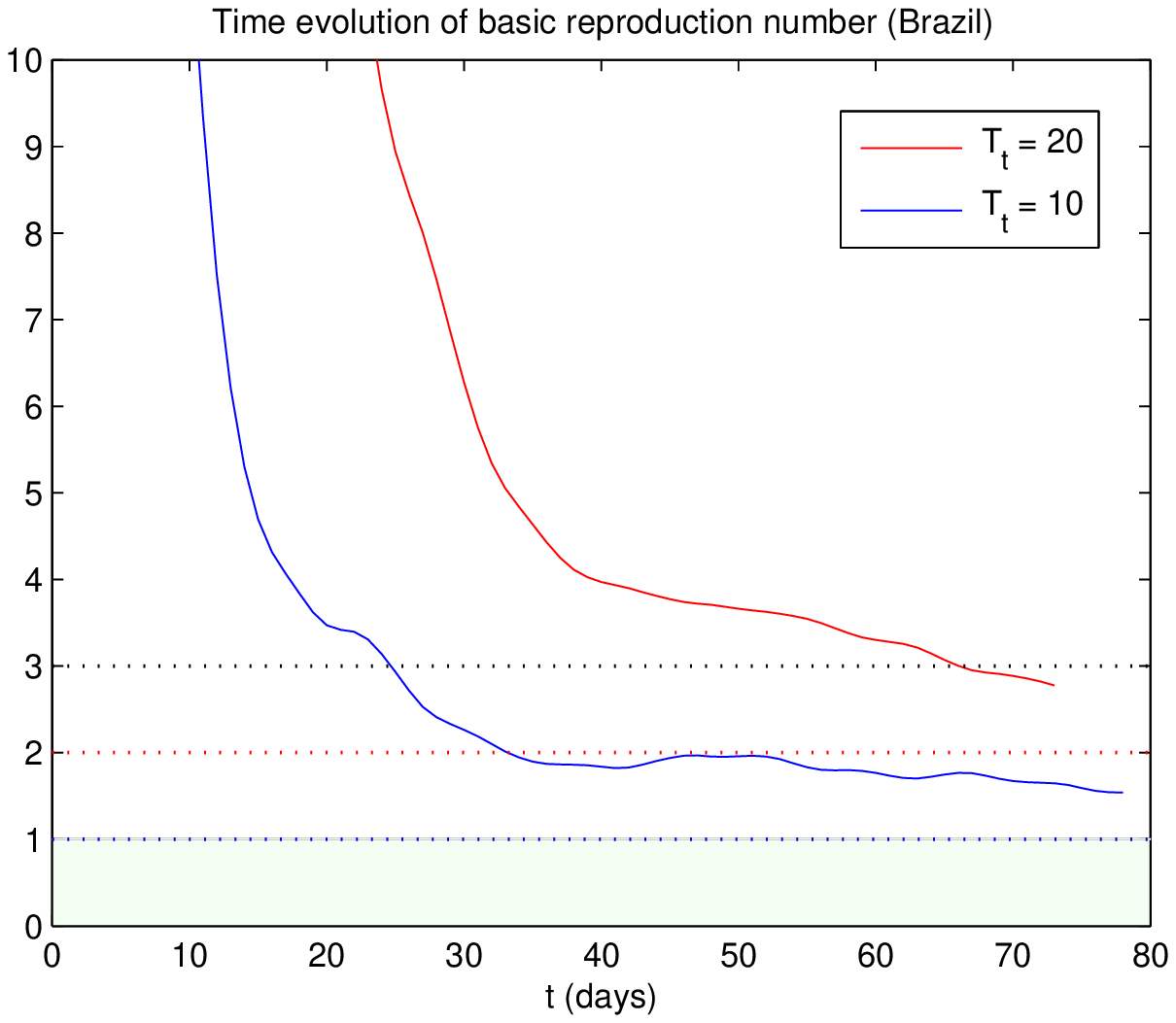}
\end{minipage}
\nl
\mbox{} \vspace{-6.750cm} \\
\mbox{} \hspace{+8.750cm}
\begin{minipage}[t]{5.350cm}
{\footnotesize {\bf Fig.\,1:}
{\footnotesize
\!Time evolution of standard \linebreak
{\em basic reproduction numbers}
of Co\:\!- \linebreak
vid-19
in Brazil
since the date of \linebreak
100 cases reported
($t = 0 $),
showing \linebreak
the effect
of two distinct hypothetical
{transmission periods}
(\:\!$T_t \!\:\!= 20$ \linebreak
\mbox{and $ T_t \!\;\!=\!\;\! 10$,
resp.).
\!In\:this\:example},
$ t = 0 $ corresponds to
03/13/2020. \\
(Data source: \href{https://covid.saude.gov.br}{covid.saude.gov.br}) \\
}
}
\end{minipage}
\nl
\mbox{} \vspace{-0.525cm} \\
\nl
%


On the other hand,
once some mathematical model has been chosen
to simulate the disease dynamics
and its parameters determined,
several alternative
reproductive numbers become
automatically available
at no additional
computational cost, \linebreak
many showing very little dependence
on key parameters like
transmission or incubation times.
We will illustrate this fact
in the context of
deterministic {\small SEIR} models,
but our approach can be adapted
to other mathematical models
(deterministic or stochastic) as well.

The idea is most easily explained
by considering the simplest {\small SEIR} model
of all, defined by the equations (1.1) below.
This model
divides the entire population in question
into four classes: the {\em susceptible\/} individuals
(class {\small S}),
those {\em exposed} (class~{\small E}, \linebreak
formed by infected people who are still inactive
(i.e., not yet transmitting the disease),
the {\em active infected} or {\em infectious\/} individuals (class {\small I})
and the {\em removed\/} ones.
%
The latter class is formed by those
who have {\em recovered\/} from the disease (class {\small R})
or who have {\em died\/} from it
(class {\small D}).
The dynamics between the various classes
is given in the universal language of calculus
by the differential equations
\newpage
\mbox{} \vspace{-0.950cm} \\
\begin{equation}
\tag{1.1}
\left\{\;
\begin{array}{l}
\mbox{$ {\displaystyle
\frac{dS}{dt} }$}
\;=\;
-\,\beta \,
\mbox{$ {\displaystyle
\frac{S(t)}{N} }$} \, I(t), \\
\mbox{} \vspace{-0.250cm} \\
\mbox{$ {\displaystyle
\frac{dE}{dt} }$}
\;=\;
\beta \,
\mbox{$ {\displaystyle
\frac{S(t)}{N} }$} \, I(t)\,-\, \delta \;\!E(t), \\
\mbox{} \vspace{-0.250cm} \\
\mbox{$ {\displaystyle
\frac{dI}{dt} }$}
\;=\;
\delta \;\!E(t) \,-\, (r + \gamma)\;\!I(t), \\
\mbox{} \vspace{-0.250cm} \\
\mbox{$ {\displaystyle
\frac{dR}{dt} }$}
\;=\;
\gamma \, I(t), \\
\mbox{} \vspace{-0.250cm} \\
\mbox{$ {\displaystyle
\frac{dD}{dt} }$}
\;=\;
r \;\! I(t),
\end{array}
\right.
\end{equation}
\nl
\mbox{} \vspace{-0.400cm} \\
see e.g.\;\cite{Brauer2008, Driessche2002, %
Hethcote2000, Martcheva2015}
for a detailed discussion of the various
terms and their meanings.
The parameters $ \beta $
({\small \sc average transmission rate}) and
$ r $ ({\small \sc average lethality rate}
of the population {\small I} due to the disease)
vary with $t$ (time, here measured in {\small \sc days}),
but $ \delta $ and $ \gamma $
are typically positive constants given by \\
\mbox{} \vspace{-0.550cm} \\
\begin{equation}
\tag{1.2}
\gamma \;=\;
\frac{1}{\,\mbox{\small $T$}_{\!\:\!t}},
\qquad
\delta \;=\;
\frac{1}{\,\mbox{\small $T$}_{\!\:\!i}},
\end{equation}
\mbox{} \vspace{-0.150cm} \\
where $\mbox{\small $T$}_{\!\:\!t}$ denotes the
{\small \sc average transmission period}
and $ \mbox{\small $T$}_{\!\:\!i}$ stands for the
{\small \sc mean incubation time},
which will be taken as 14 and 5.2, respectively
\cite{Kucharski2020, Lauer2020, Vaid2020}).
\!In the system (1.1),
$N$ denotes the full size of the susceptible population
initially exposed,
so that we have
$ S(t_0) + E(t_0) + I(t_0) $
$ + R(t_0) + D(t_0) = N $,
where $t_0$ denotes the initial time.
\!Observing that, by the equations (1.1),
the sum $ \:\!S(t) + E(t) + I(t) + R(t) $
$ + D(t) \:\!$
is invariant,
it follows the {\small \sc conservation law} \\
\mbox{} \vspace{-0.550cm} \\
\begin{equation}
\tag{1.3}
S(t) + E(t) + I(t) + R(t) + D(t) \;\!=\, N,
\qquad
\forall \;\,
t > t_0,
\end{equation}
\mbox{} \vspace{-0.200cm} \\
since, for simplicity,
the model neglects any
changes in the population
due to birth, migration
or death by other causes
during the period of the epidemic
(of the order of a few months).
To well define the model (1.1),
besides informing the functions
$ \beta(t) $ and $r(t)$
we need to provide
the initial values
$ S(t_0) $, $ E(t_0) $, $ I(t_0) $, $ R(t_0) $, $ D(t_0) $,
which is not a trivial task,
since not all of these variables
are reported,
and those reported may be in error
--- which may well be large
in case of significant underreporting.

It thus seems clear that
predicting reasonably right values
for the variables
$S(t)$, $ E(t)$, $I(t)$,
$R(t)$ and $ D(t) $
at future times
is {\em not\/} a simple problem,
especially in the long time range.
The situation becomes
even more complicated
for more complex (i.e., stratified) models,
which add other
variables and parameters
to be determined.
Calibrating many parameters
can quickly become a nightmare.
For all its simplicity,
models with few variables and parameters
like (1.1)
can yield surprisingly good results
and thus should not be overlooked,
as will be seen in the sequel.
\newpage
%

%
%
%
%

%
\mbox{} \vspace{-1.250cm} \\
{\bf 2. Implementing the SEIR model} \\
%

Having introduced the {\small SEIR} equations (1.1),
we now describe an implementation of this model
that is suitable for the computation of
reproduction numbers. \\
\mbox{} \vspace{-0.700cm} \\

({\em i\/}) {\em assigning a value to
the population parameter $\!\;\!N$} \\
\mbox{} \vspace{-0.750cm} \\

In the case of {\small C}ovid-19,
which can be considered a new virus
({\small SARS}-{\small C}o{\small V}-2),
it has been common to assume
the entire population susceptible
and assign its whole value to $N\!\;\!$.
This is highly debatable,
since this parameter refers to that
particular fraction \linebreak
of the susceptible population
that is effectively subject to infection.
\!For deterministic models,
this introduces the possibility
that an outbreak might {\em not\/} happen
after the introduction or reintroduction
of a few infected individuals,
as it has been long recognized in the
stochastic literature
\cite{Allen2008, Kucharski2020}.
In any case, it turns out that
$N$ is not so much important
for the short range dynamics
as it proves to be
in the long run (see Figures 2$a$ and 2$b$),
so that for our present purposes
this is not a serious issue.
We have therefore taken for
$N$ the full population
of the region under consideration. \\
\nl
\mbox{} \vspace{-0.250cm} \\
%
%
%
%
%
%
\mbox{} \hspace{+1.250cm}  
\begin{minipage}[t]{14.600cm}  
\includegraphics[keepaspectratio=true, scale=0.80]{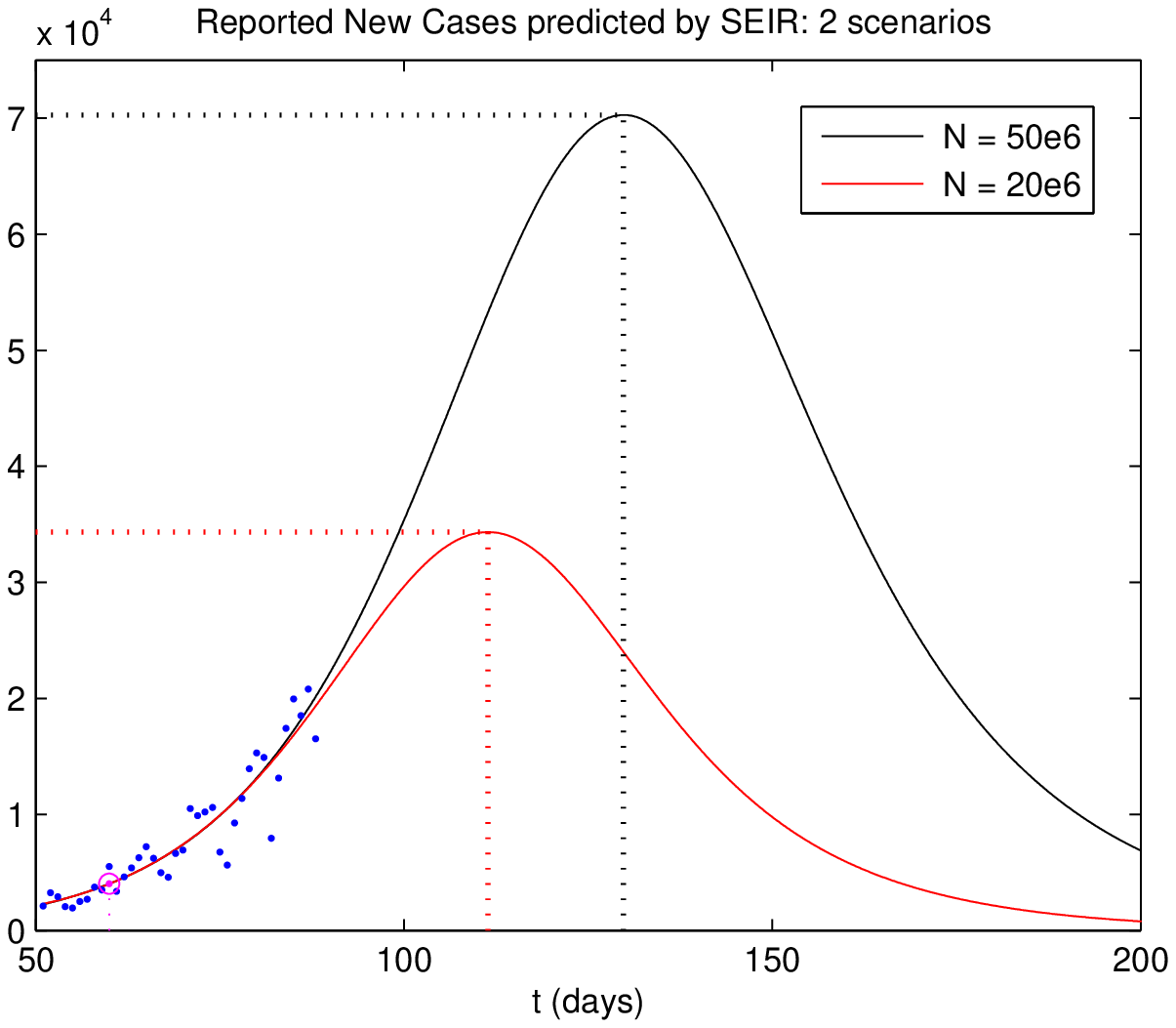} 
\end{minipage} \\
\mbox{} \vspace{-0.350cm} \\
\mbox{} \hspace{+0.250cm}
{\footnotesize {\bf Fig.\,2\!\;\!{\em a\/}:}
\begin{minipage}[t]{12.200cm}
{\footnotesize
Prediction by model (1.1)
of the daily number of {\em new cases\/} of Covid-19
expected to be reported in Brazil
between the initial time
$ t = t_0 = 60 $ (April\;25th)
and $ t = 200 $ (September\;\!\;\!12th),
considering susceptible populations
of $N \!\:\!= 20 $ million (red curve)
and $ N \!\:\!= 50 $ million (black curve).
Note the appreciable difference between \linebreak
the predicted peak values
(34 and 70 thousand, resp.)
\!\:\!and their respective dates, \linebreak
June\:6th and July\:4th.
Actual data points are shown in blue.
(\:\!Computed from
data available at the official site\;\!
\url{https://covid.saude.gov.br}.) \\
}
\end{minipage}
}
\newpage
\mbox{} \vspace{-0.950cm} \\
%
%
%
%
%
\mbox{} \hspace{+1.250cm}  
\begin{minipage}[t]{14.600cm}  
\includegraphics[keepaspectratio=true, scale=0.80]{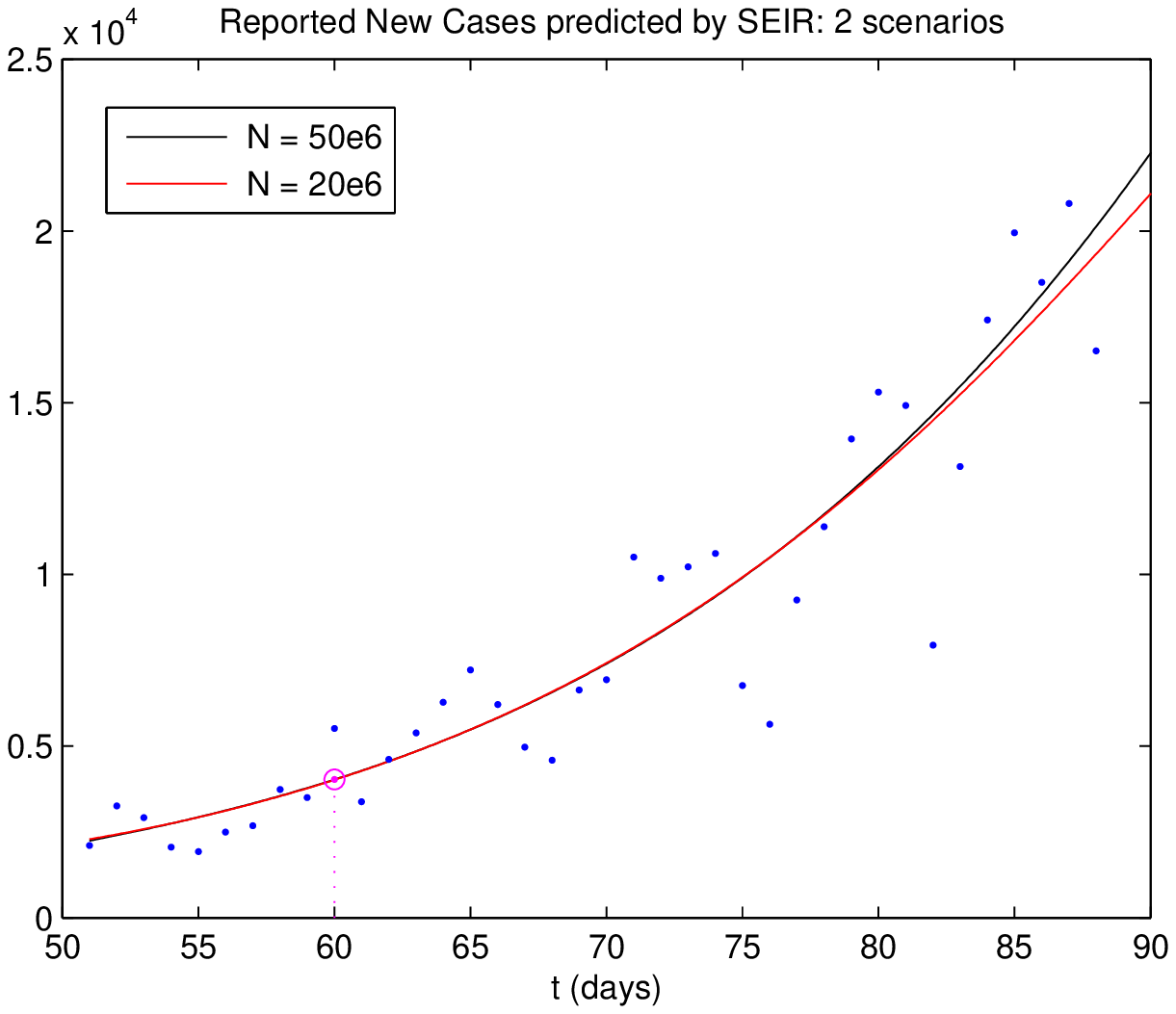}
\end{minipage}
\mbox{} \vspace{-0.350cm} \\
\mbox{} \hspace{+0.350cm}
{\footnotesize {\bf Fig.\,2{\em b\/}:}
\begin{minipage}[t]{12.500cm}
{\footnotesize
Thirty day prediction by model (1.1)
of the daily number of {\em new cases\/} of Covid-19
to be reported in Brazil
between the initial time
$ t = t_0 = 60 $ (04/25)
and $ t = 90 $ (05/25),
considering susceptible exposed populations
of $N \!\;\!= 20 $ million (red curve) \linebreak
and $ N \!\;\!= 50 $ million (black curve).
Note the very close similarity of the two
30D predictions in spite of the appreciable
difference in the values of $N\!\:\!$.
Points shown in blue are the official values
reported (cf.\url{https://covid.saude.gov.br}.)
}
\end{minipage}
\mbox{} \hspace{+1.000cm} 
}
\mbox{} \vspace{-0.500cm} \\

({\em ii\/}) {\em generation of initial data\/}
$S(t_0)$, $ E(t_0) $, $ I(t_0)$, $ R(t_0)$, $ D(t_0)$ \\
\mbox{} \vspace{-0.750cm} \\

Initial values
$S_0, E_0, I_0, R_0, D_0 $
for the five variables are generated
from a starting date $ t_{\!\;\!s}\!\;\!$ on,
which is taken so as to meet some minimum value
chosen
of total reported cases (typically, 100).
\!Denoting by $ C_{r}(t) $
the total amount of reported cases
up to some time $t$,
and letting \mbox{\small EIR}$(t)$
be the sum of the populations
$ E(t) $, $ I(t) $ and $ R(t) $,
we set \\
\mbox{} \vspace{-0.950cm} \\
\begin{equation}
\tag{2.1}
\mbox{\small EIR}(t_s) \,=\,
f_{c} \!\;\!\cdot (\:\!C_{r}(t_s) - D(t_s)),
\end{equation}
\mbox{} \vspace{-0.275cm} \\
where $f_{c} \!\;\!\geq\!\;\! 1 $
denotes a {\small \sc correction factor}
to account for likely underreportings
on the official numbers given.
(\:\!In (2.1), we have neglected possible
underreportings on the number of deaths,
which could of course be similarly accounted for
if desired.)
Again, this factor will not play an important role
in this paper and could be safely ignored,
but it should be carefully considered
in the case of long time predictions.
Having estimated $ \mbox{\small EIR}(t_s) $,
we then set \\
\mbox{} \vspace{+0.000cm} \\
\mbox{} \hspace{+2.250cm}
$ {\displaystyle
E(t_s) \,=\, E_0(t_s)
:=\; a \cdot (1 - b) \!\:\!\cdot \mbox{\small EIR}(t_s)
} $,
\hfill (2.2$a$) \\
\mbox{} \vspace{-0.150cm} \\
\mbox{} \hspace{+2.250cm}
$ {\displaystyle
I(t_s) \,=\, I_{0}(t_s)
:=\; (1 - a) \cdot (1 - b) \!\:\!\cdot \mbox{\small EIR}(t_s)
} $,
\hfill (2.2$b$)
\newpage
\mbox{} \vspace{-0.750cm} \\
\mbox{} \hspace{+2.250cm}
$ {\displaystyle
R(t_s) \,=\, R_{0}(t_s)
:=\; b \cdot \mbox{\small EIR}(t_s)
} $,
\hfill (2.2$c$) \\
\mbox{} \vspace{-0.100cm} \\
\mbox{} \hspace{+2.250cm}
$ {\displaystyle
S(t_s) \,=\,  S_{0}(t_s) :=\,
N -\, \bigl(\;\! E(t_s) + I(t_s) + R(t_s) + D(t_s) \:\!\bigr)
} $,
\hfill (2.2$d$) \\
\mbox{} \vspace{-0.050cm} \\
where $ \!\;\!\;\!a =\!\;\!\;\! \mbox{\small $T$}_{\!\:\!i} /
(\mbox{\small $T$}_{\!\:\!i} +\!\;\!\;\!\mbox{\small $T$}_{\!\:\!t}) $
and $\!\;\!\;\! b \!\;\!\;\!=\!\;\!\;\! 0.30 $,
consistently with the literature
(see e.g.\;\cite{Vaid2020}). \\
The arbitrariness in this choice of weights
gets eventually corrected
as we compute more values
$ S_0(t_0), E_0(t_0), I_0(t_0), R_0(t_0), D_0(t_0) $
at later initial times
\mbox{$ \:\!t_0  \!\:\! = t_s \!\:\!+\!\:\! 1,\!\:\!..., t_{\mbox{}_{\!\:\!F}} \!\;\!$},
where \mbox{$ \:\! t_{\mbox{}_{\!\:\!F}} \!\:\!$}
stands for the final (i.e., most recent)
date of reported data available.
For each $ t_0 $,
the solution of the equations (1.1)
with the previously obtained initial data
at \mbox{\:\!$ t_0 \!\:\!-\!\:\! 1 $}
is computed on the interval
\mbox{$ J(t_0) = [\,t_0\!\:\! -\!\:\! 1,\;\! t_1\;\!] $},
\mbox{$ t_1 \!\:\!=\;\! \min\;\!
\{\;\! t_0\!\:\! -\!\:\! 1\!\;\! + d_0, \, t_{\mbox{}_{\!\:\!F}} \!\;\!\} $},
with constant parameters
\mbox{$ \beta = \beta_0(t_0 \!\:\!-\!\:\! 1)$},
\mbox{$ r = r_0(t_0 \!\:\!-\!\:\! 1)$}
determined so that
the com\:\!- \linebreak
puted values for
$C_r(t) $, $ D(t) $
best fit the reported data
for these variables
on \mbox{$ [\;\!t_0, \;\!t_{1}\:\! ] $}
in the sense of {\small \sc least\;squares}
\cite{Martcheva2015}.
\!(\:\!Here,
\mbox{$ d_0 \!\;\!\in [\,2, \;\!10\;\!] $}
is chosen according to the data regularity.)
Once this solution $(S,E,I,R,D)(t)$
is obtained,
we set
$ S_0(t_0) \!:= S(t_0) $,
$ E_0(t_0) \!:= E(t_0) $,
$ I_0(t_0) \!:= I(t_0) $,
$ R_0(t_0) \!:= R(t_0) $,
$ D_0(t_0) \!:= D(t_0) $
and move on to the
next time level
\mbox{$\:\!t_0 \!\:\!+\!\:\! 1$},
repeating the procedure until
\mbox{$t_{\mbox{}_{\!\:\!F}}$}\!\:\!
is reached. \\
\mbox{} \vspace{-0.550cm} \\

({\em iii\/}) {\em computing the solution on
some final interval} $ [\,t_0, \;\!\mbox{\small $T$}\;\!] $
\!({\small \sc prediction phase}) \\
\mbox{} \vspace{-0.650cm} \\

Having completed the previous steps,
we can address the possibility of
{\em prediction}.
Although this is not important for our present goals,
it is included for completeness.
Choosing an initial time
\mbox{$\:\!t_0 \!\;\!\in (\:\! t_{s}, \;\!t_{\mbox{}_{\!\:\!F}}] $},
we then take the initial values \\
\mbox{} \vspace{-0.650cm} \\
\begin{equation}
\notag
S(t_0) = S_{0}(t_0),
\;
E(t_0) = E_{0}(t_0),
\;
I(t_0) = I_{0}(t_0),
\;
R(t_0) = R_{0}(t_0),
\;
D(t_0) = D_{0}(t_0).
\end{equation}
\mbox{} \vspace{-0.200cm} \\
In order to predict the values of
the variables
$ S(t), E(t), I(t), R(t), D(t) $
for $ t > t_0 $,
it is important to have
good estimates for the
evolution of the key parameters
$ \beta(t) $ and $ r(t) $
beyond $t_0$.
This is the most computationally intensive
part of the algorithm
and is better executed
in large computers.
\!Such estimates can be
given in the form \\
\mbox{} \vspace{-0.550cm} \\
\begin{equation}
\tag{2.3$a$}
\beta(t) \,=\; \beta_0 \;\!+\, a_{\beta} \,
e^{\mbox{\footnotesize $ -\,\lambda_{\beta}(\:\!t - \:\!t_0)$}}
\end{equation}
\mbox{} \vspace{-0.900cm} \\
\begin{equation}
\tag{2.3$b$}
r(t) \,=\; r_0 \;\!+\, a_{r} \,
e^{\mbox{\footnotesize $ -\,\lambda_{r}(\:\!t - \:\!t_0)$}}
\end{equation}
\mbox{} \vspace{-0.225cm} \\
where
$\beta_0, a_{\beta}, \lambda_{\beta}, r_0, a_{r}, \lambda_{r}
\!\;\!\in \mathbb{R} $
are determined
so as to minimize the maximum size
of weighted {\small \sc relative errors\/}
in the computed values for
$C_{r}(t), D(t) $
as compared to the official data reported
for these variables
on some previous interval
\mbox{$ [\;\! t_{0}\!\:\!-\!\:\!\tau_{0}, \;\!t_{0} \:\!] $}
(weighted {\small \sc Chebycheff} {\small \sc problem})
for some chosen
$ \tau_0 > 0 $
(usually, $ 20 \leq \tau_0 \leq 30 $).
This problem is solved iteratively
starting with an initial guess
obtained from the analysis of the
previous values $ \beta_0(t), r_0(t) $
computed in the step ({\em ii\/}) above.
\!\mbox{The result} is illustrated
in Figure\;3 for the case of $ \beta(t) $,
with similar considerations for $ r(t)$.
\newpage
\mbox{} \vspace{-1.500cm} \\
%
%
%
%
%
\mbox{} \vspace{-0.400cm} \\
\mbox{} \hspace{+1.500cm}
\begin{minipage}[t]{14.600cm}  
\includegraphics[keepaspectratio=true, scale=0.75]{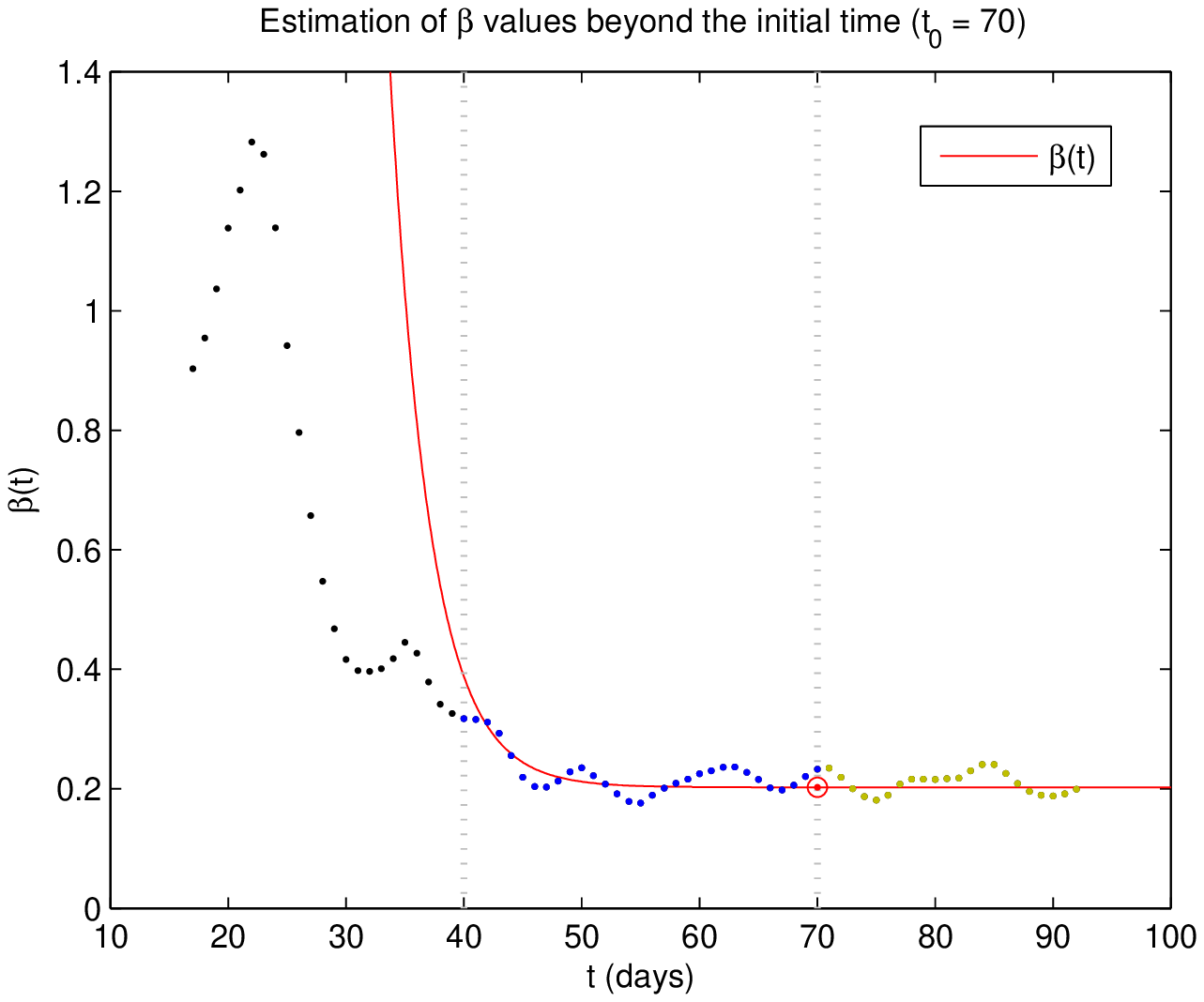} 
\end{minipage}
\mbox{} \vspace{-0.650cm} \\
\mbox{} \hspace{-0.250cm}
{\footnotesize {\bf Fig.\,3:}
\begin{minipage}[t]{14cm} 
{\footnotesize
Estimation of future values of
the transmission parameter $ \beta(t) $
beyond the initial time $ t_{0} $
$ = 70 $ (05/05/2020)
for the outbreak of Covid-19 in Brazil,
assuming the basic form (2.3$a$),
after solving the
Chebycheff problem (red curve).
The data points in the interval
\mbox{$ [\;\!40, 70\;\!] $},
shown here in blue,
are values of the function $ \beta_0(t) $
computed in step ({\em ii\/}),
which are used to obtain the
first approximation to $ \beta(t) $.
Values of $ \beta_0(t)$
previous to $ t = 40 $ (04/05/2020),
shown in black, are disregarded.
\!The golden points beyond $ t_0 \!\;\!= 70 $
are future values of $ \beta_0(t) $, \linebreak
not available on 05/05/2020,
displayed to allow comparison
with the predicted values $ \beta(t) $. \linebreak
}
\end{minipage}
}
\nl
\mbox{} \vspace{-0.500cm} \\
Once $ \beta(t) $, $ r(t) $
have been obtained,
the equations (1.1) are finally solved
(Figure 4). \linebreak
\mbox{} \vspace{-0.550cm} \\
%

%
%
%
%
\mbox{} \vspace{-0.600cm} \\
\mbox{} \hspace{+1.500cm}
\begin{minipage}[t]{14.600cm}  
\includegraphics[keepaspectratio=true, scale=0.80]{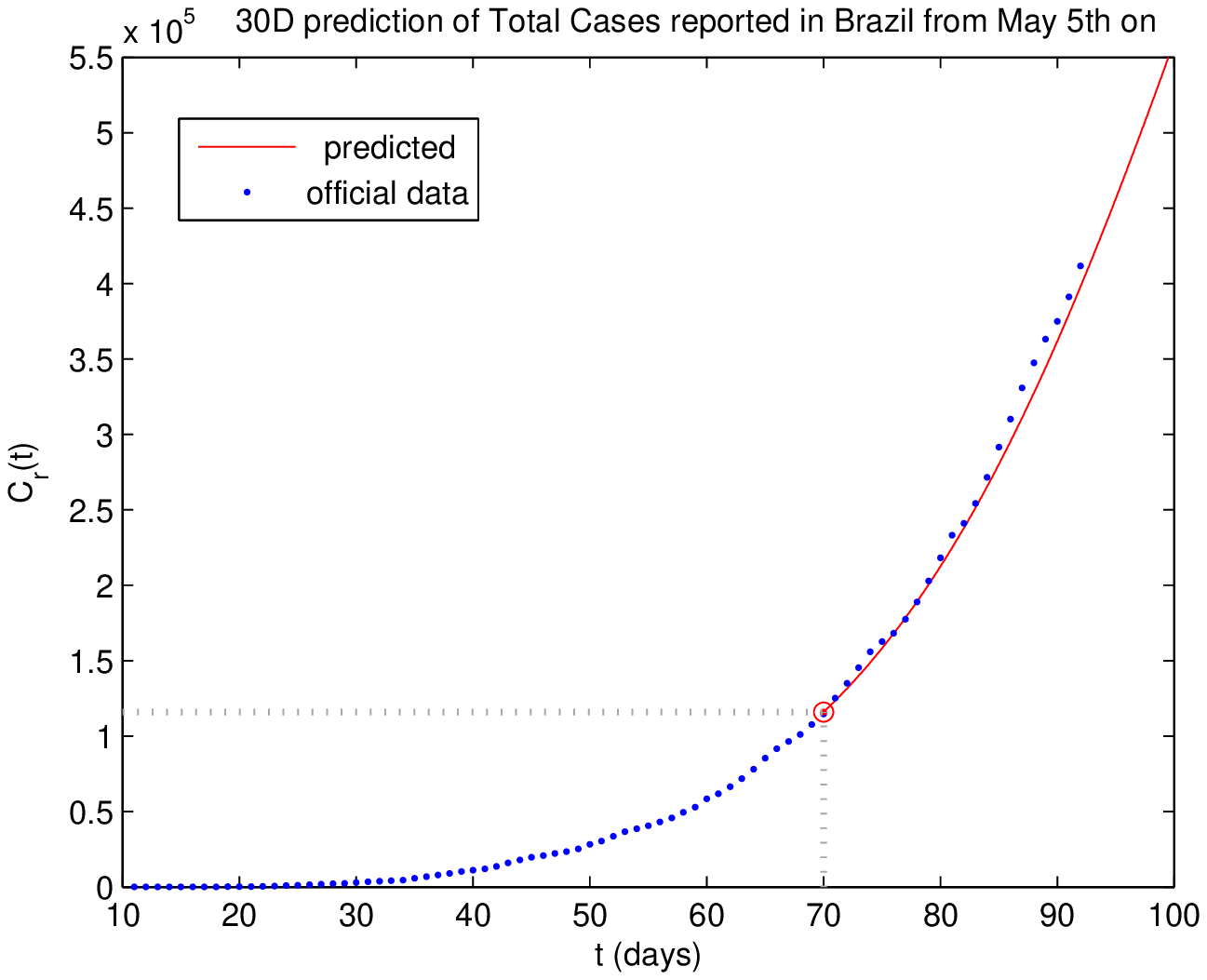} 
\end{minipage}
\mbox{} \vspace{-0.600cm} \\
\mbox{} \hspace{-0.250cm}
\nopagebreak
{\footnotesize {\bf Fig.\,4:}
\begin{minipage}[t]{14cm}
{\footnotesize
Computation of
$ C_{r}(t) = \bigl( E(t) + I(t) + R(t) \bigr)/f_{c}
+ D(t) $ for $\;\! t > t_0 = 70 $ (05/05/2020),
with initial data
$ C_{r}(t_0) = \bigl(E_0(t_0) + I_0(t_0) + R_0(t_0) \bigr)/f_{c} + D_0(t_0) $,
after obtaining $ \beta(t) $, $ r(t) $ -- see Fig.\,3 for $ \beta(t)$.
The numerical solution of equations (1.1)
is easily obtained by any method.
}
\end{minipage}
}
\newpage
\mbox{} \vspace{-0.950cm} \\
%
%
%
%

{\bf 3. Reproduction numbers} \\

A natural by-product of the results
generated by the algorithm is the
estimate of {\em reproduction numbers\/}
of the epidemic,
which measure the intensity of transmission
at various times
and, in doing so,
are useful indicators
to monitor the situation
and the effects of
intervention procedures
that may have been taken.
Using the generic symbol
$ \!\;\!R_{t} $
to denote such quantities,\!\footnote{%
%
%
%
%
The notation $R_{t}$ is natural
in stochastic models,
and is adopted here
as we have already
used $ R(t) $, $ R_0(t) $
with other meanings
(size of the recovered population
and their initial values,
resp.).
}
%
%
they signal a rise in the number of infections
in the case $ \!\;\!R_t \!\;\!> 1 $,
their decrease when
$ \!\;\!R_t \!\;\!< 1 $,
and
temporary steadiness
if $ \!\;\!R_t \!\;\!= 1 $.
For instance,
rewriting the equation for
the critical population $I(t)$
in the form \\
\mbox{} \vspace{-0.550cm} \\
\begin{equation}
\tag{3.1$a$}
\frac{d\,\!I}{d\:\!t}
\;=\;
\alpha(t) \;\!I(t),
\qquad \;\,
\alpha(t) :=\, \delta \!\:\!\cdot\!\:\! E(t)\:\!/\:\!I(t)
\;\!-\;\!r(t) -\:\! \gamma,
\end{equation}
\mbox{} \vspace{-0.150cm} \\
we see that $ I(t) $ will increase
if $ \alpha(t) > 0 $,
decrease when $ \alpha(t) < 0 $
and
stay about the same if $ \alpha(t) = 0 $
--- or, in terms of
the ratio \\
\mbox{} \vspace{-0.550cm} \\
\begin{equation}
\tag{3.1$b$}
R_{t}
\!\;\! :=\;
\frac{\;\!\delta \!\:\!\cdot\!\:\! E(t)\:\!/\:\!I(t)\:\!}
{r(t) + \gamma},
\end{equation}
\mbox{} \vspace{-0.150cm} \\
whether we have
$ R_{t} \!\;\!> 1 $,
$ R_{t} \!\;\!< 1 $
or $ R_{t} \!\;\!= 1 $,
respectively.
Another natural possibility
would be to consider
basic ratios like \\
\mbox{} \vspace{-0.550cm} \\
\begin{equation}
\tag{3.2}
R_{t}
\!\;\! :=\;
\frac{\;\!I(t+d)\:\!}{\;\!I(t-d)\:\!},
\qquad \;\;
R_{t}
\!\;\! :=\;
\frac{\;\!E(t+d) + I(t+d)\:\!}
{\;\!E(t-d) + I(t-d)\:\!}
\end{equation}
\mbox{} \vspace{-0.100cm} \\
for some chosen $d \!\;\!>\!\;\! 0 $.
For example,
the choice $ d = \mbox{\small $T$}_{\!\:\!t}\mbox{\small $/2$} $
corresponds to the standard
{\em basic reproduction number},
or the mean number of secondary infections
caused by a typical infected individual
during his transmission period
\cite{Kucharski2020, Martcheva2015}.
The corresponding expressions
would be, using the calculations
performed in step ({\em ii\/})
of the algorithm, \\
\mbox{} \vspace{-0.500cm} \\
\begin{equation}
\tag{3.3}
R_{t}^{\;\!(1)}
\!\;\! :=\;
\frac{\;\!\delta \!\:\!\cdot\!\:\! E_0(t)\:\!/\:\!I_0(t)\:\!}
{r_0(t) + \gamma},
\end{equation}
\mbox{} \vspace{-0.150cm} \\
where $ r_0(t) $ denotes
the lethality rates computed there,
or else \\
\mbox{} \vspace{-0.500cm} \\
\begin{equation}
\tag{3.4}
R_{t}^{\;\!(2)}
\!\;\! :=\;
\frac{\;\!I_0(t+3)\:\!}{\;\!I_0(t-3)\:\!},
\qquad \;\;
R_{t}^{\;\!(3)}
\!\;\! :=\;
\frac{\;\!E_0(t+3) + I_0(t+3)\:\!}{\;\!E_0(t-3) + I_0(t-3)\:\!},
\end{equation}
\mbox{} \vspace{-0.100cm} \\
and so forth.
These indicators
point to similar scenarios
(Figura\,5),
%
%
with
$ R_{t}^{\;\!(1)} \!$
seemingly more influenced
by seasonal (weekly) variations in the data.
\!We have found $ R_{t}^{\;\!(2)} \!\:\!$
particularly useful,
with numerical results
that are consistent with previous analyses
\linebreak
(see e.g.\;\cite{Mellan2020}).
For time scales such as
those of {\small C}ovid-19,
the choice $ d = 3 $
is good to
zoom in the scenario
and facilitate the reading
(Figure 6),
while not compromising robustness
(Figure 7). \\

\mbox{} \vspace{-0.850cm} \\
%

%
%
%
%

%
\mbox{} \hspace{+1.500cm}
\begin{minipage}[t]{15.00cm}
\includegraphics[width = 10.00cm, height = 7.50cm]{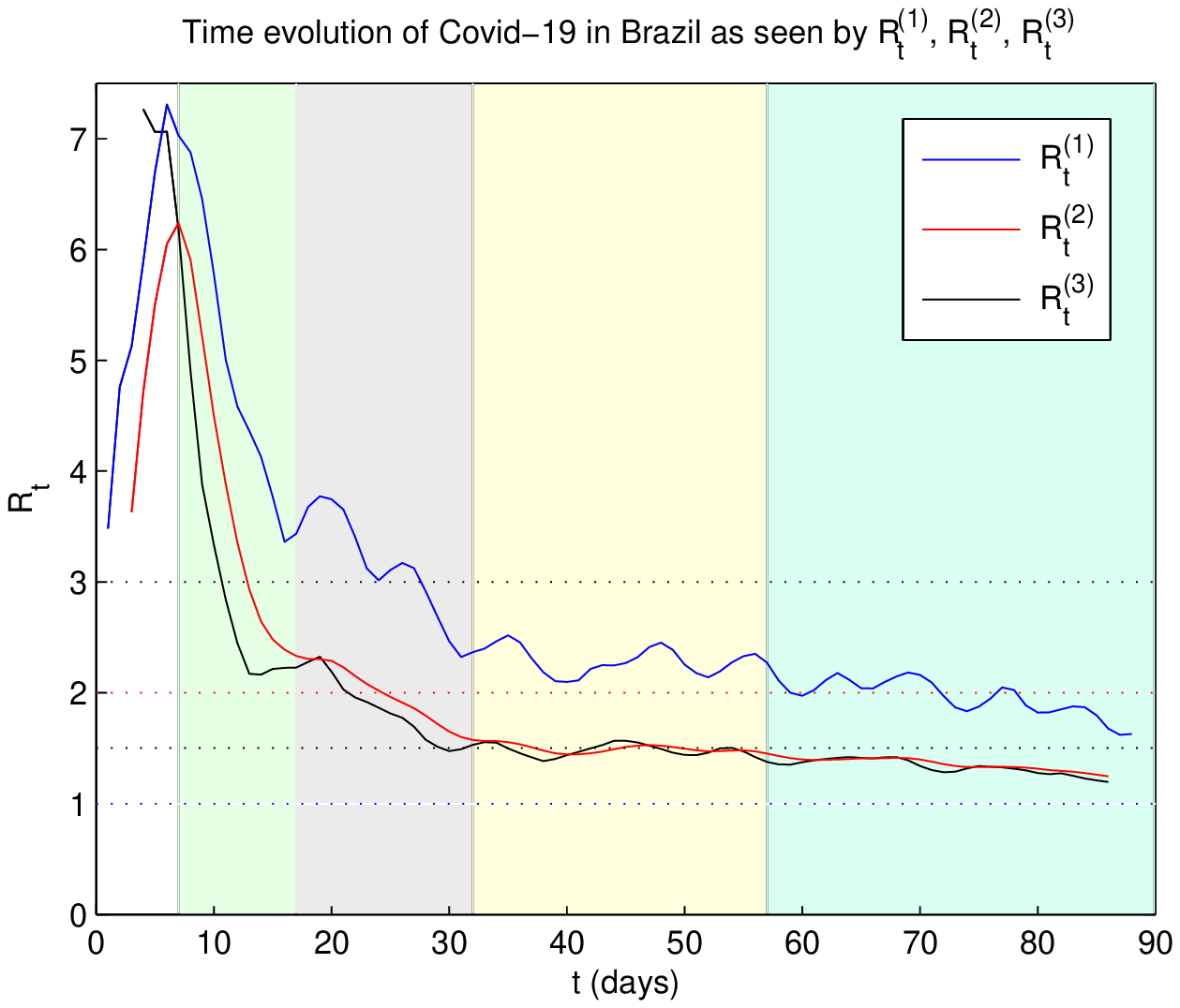}
\end{minipage}
\mbox{} \vspace{-0.600cm} \\
\mbox{} \hspace{+0.025cm}
{\footnotesize {\bf Fig.\,5:}
\begin{minipage}[t]{13.750cm}
{\footnotesize
Comparison of the time evolution of Covid-19
in Brazil (since 100 cases reported)
as seen by the indicators defined
in (3.3), (3.4),
pointing to similar scenarios.
%
%
In the three cases it is clear
that Brazil has not yet reached a
state of control of the epidemic
($R_t \!\;\!<\!\:\!1 $)
}
\end{minipage}
}
\mbox{} \vspace{+0.500cm} \\
%

%
%
%
%

%
\mbox{} \hspace{+1.500cm}
\begin{minipage}[t]{15.00cm}
\includegraphics[width = 10.00cm, height = 7.50cm]{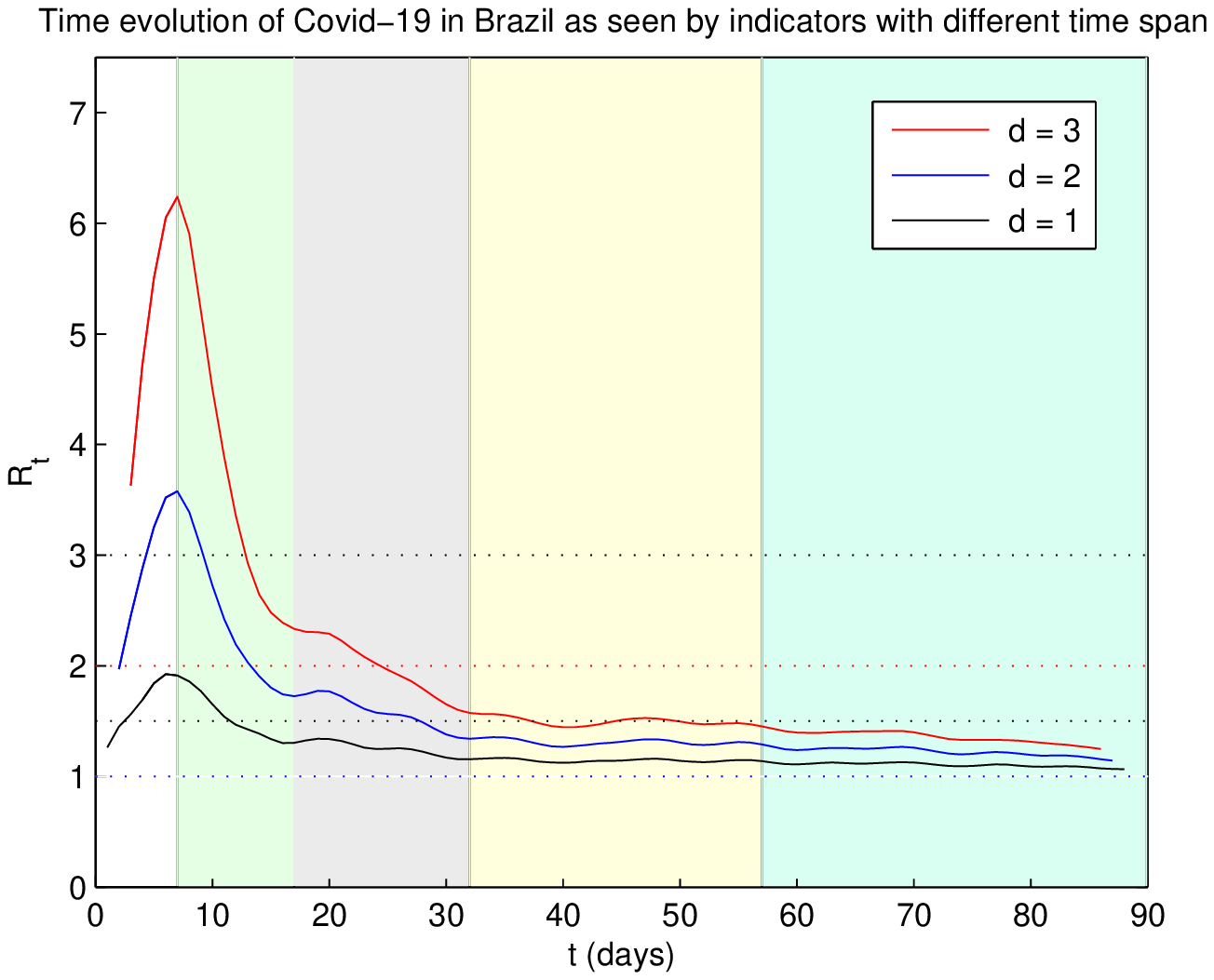}
\end{minipage}
\mbox{} \vspace{-0.600cm} \\
\mbox{} \hspace{+0.025cm}
{\footnotesize {\bf Fig.\,6:}
\begin{minipage}[t]{13.750cm}
{\footnotesize
Comparison of the time evolution of Covid-19
in Brazil (since 100 cases reported)
as seen by
$ R_t = I(t+d)/I(t-d) $
for different values of $d$,
showing similar scenarios.
%
%
In the three cases it is clear
that Brazil has not yet reached a
state of control of the epidemic
($R_t \!\;\!<\!\:\!1 $)
}
\end{minipage}
}
\mbox{} \vspace{-0.300cm} \\
\newpage
\mbox{} \vspace{-1.250cm} \\
%
%
%
%
%
%

%
\mbox{} \hspace{+1.500cm}
\begin{minipage}[t]{15.00cm}
\includegraphics[width = 10.00cm, height = 7.50cm]{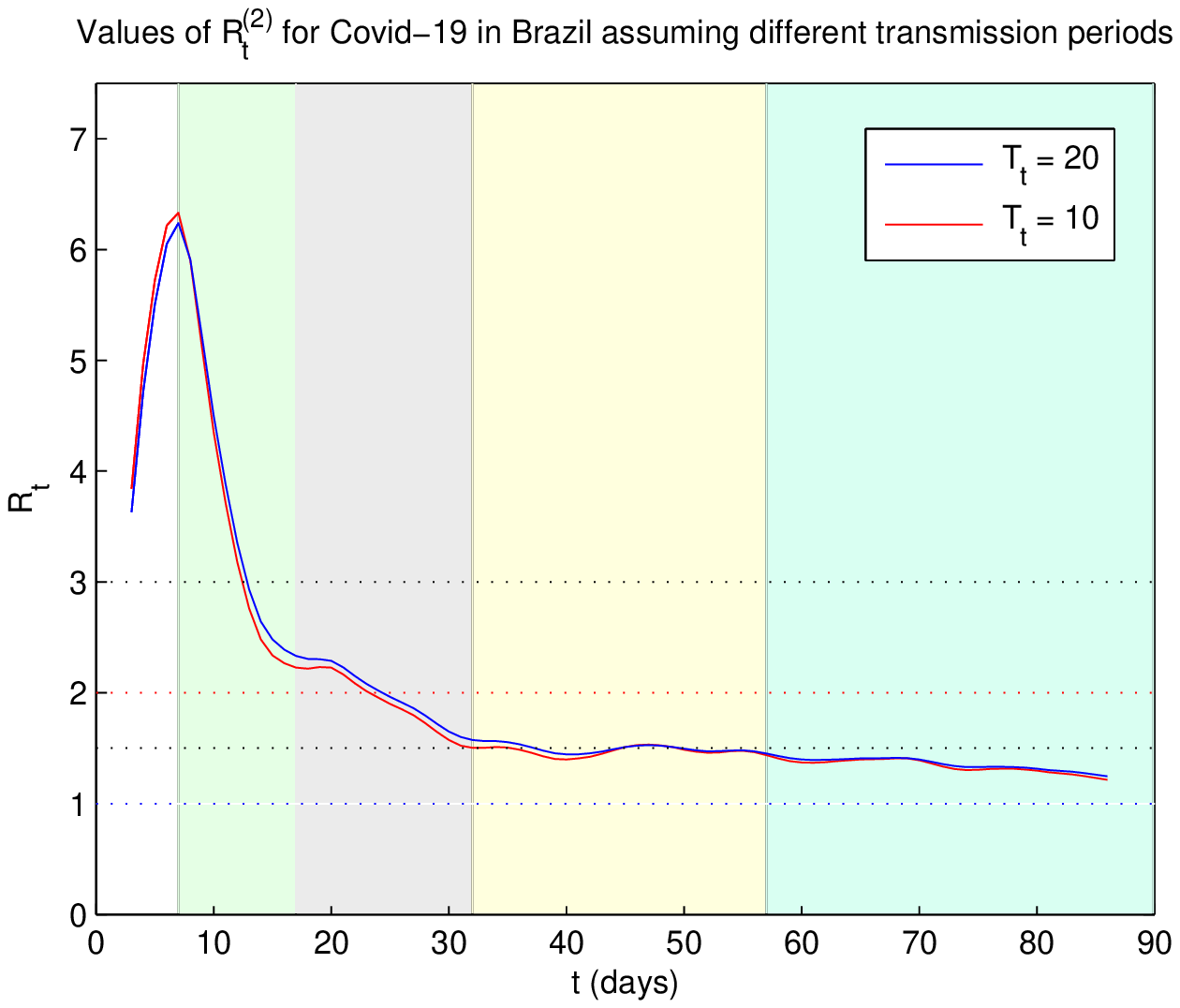}
\end{minipage}
\mbox{} \vspace{-0.750cm} \\
\mbox{} \hspace{+0.025cm}
{\footnotesize {\bf Fig.\,7:}
\begin{minipage}[t]{13.500cm}
{\footnotesize
Robustness of $ R_t^{(2)} \!$
with respect to large uncertainties
on the value of transmission time. \linebreak
Date zero refers to 100 cases reported,
that is: 03/13/2020.
\!(As in Fig.\,5 and Fig.\,6 above, \linebreak
calculations were based upon official data
reported at\;\! 
\url{https://covid.saude.gov.br}.)  \\
%
%
}
\end{minipage}
}
\mbox{} \vspace{-0.050cm} \\
%
%
%
%

{\bf 4. Applications} \\

In this section we will illustrate
the use of reproduction values
by examining the evolution
of Covid-19 in various countries
around the world 
under the view of 
such numbers --- 
choosing for definiteness
the numeric ratio $R_t^{(2)}\!$
defined in (3.4) above
as our basic indicator,
unless explicitly stated otherwise.
Thus,
we set \\
\mbox{} \vspace{-0.500cm} \\
\begin{equation}
\tag{4.1}
R_{t} \,=\;
\frac{\;\!I_0(t+3)\:\!}{\;\!I_0(t-3)\:\!}
\end{equation}
\mbox{} \vspace{-0.100cm} \\
where $ I_0(s) $
is the size of the active infected population
at time $s$ as computed
in the step ({\em ii\/})
of the {\small SEIR} algorithm
(see Section 2). \\
\mbox{} \vspace{-0.750cm} \\

Taking right decisions about intervention
or relaxation measures is a very difficult
and complex process that involves
a careful consideration of
various mathematical indicators and 
a lot of other factors including many health,
economic and social issues.
In the following examples
we consider only the single factor
given by reproduction numbers. 
For all the simplicity and obvious limitations
of this approach,
it offers nevertheless precious insight and
information about the disease dynamics and evolution. \\
\nl
{\bf Acknowledgements.}
In the following examples,
the computation of 
the $R_t$ curves \linebreak
was based on data
available for each country
at the site 
\href{https://www.worldometers.info/coronavirus/}{\color{blue} worldometers/coronavirus}. \linebreak
%

%
%
%
%
\mbox{} \hfill 
\begin{minipage}[t]{9.10cm}  
\includegraphics[keepaspectratio=true, scale=0.70]{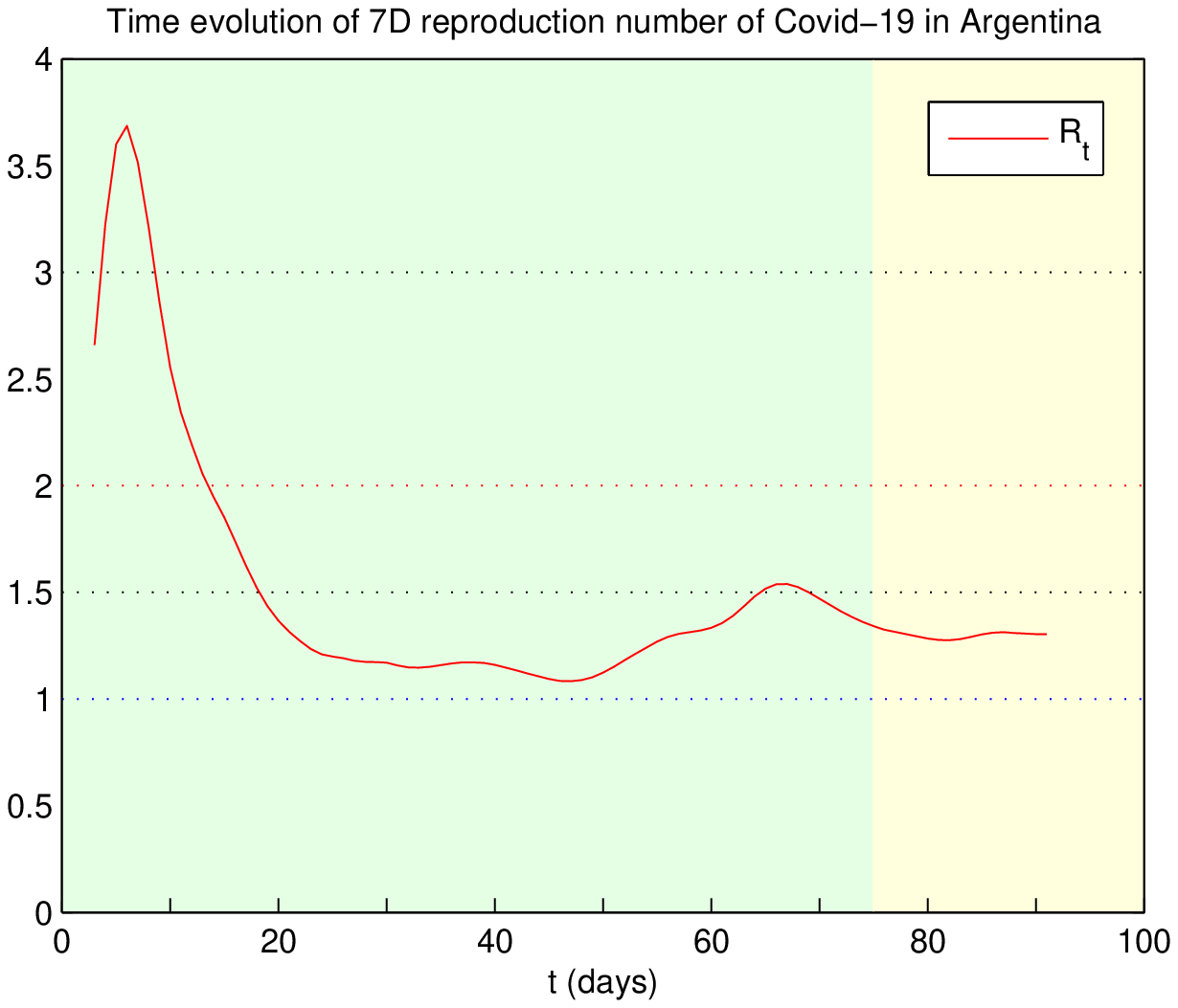}
\end{minipage}
\mbox{} \vspace{-7.850cm} \\
\mbox{} \hspace{-0.250cm}
\begin{minipage}[t]{6.000cm}
{\small \bf Example 1:}
{\small
Time evolution of Co\:\!- \linebreak
vid-19 in Argentina since 03/18/ \linebreak
2020 ($ t = 0 $), the date of
97 total cases reported.
\!Strong containment \linebreak
measures
had begun 3 days earlier
($t = -\,3$)
and managed to keep
the number of cases and deaths
down low,
with $R_t$ decreasing
continually until
05/04/2020 ($ t = 47$),
when it reached
a minimum value of 1.08. \linebreak
Following that,
the situation deteriorated
with $R_t$ increasing
to 1.54 \linebreak
on 05/24/2020
($ t = 67 $),
despite the reinforcement
of most intervention \linebreak
}
\end{minipage}
\mbox{} \vspace{-0.325cm} \\
{\small
procedures.
\!\!\;\!Partial relaxation of some of these measures
was introduced on 06\!\;\!/01\!\:\!/\!\;\!2020 ($ t = \!\;\!75 $)
and,
in this new period,
\!\mbox{$ R_t \!\:\!$ has\:remained\:relatively\:stable
at \!\:\!1.30 \!(yellow\:band).}
}
Bringing the epidemic to a state of
nationwide control
($R_t \!\;\!<\!\;\!1 $)
still seems far away.
\!This ex\-am\-ple illustrates
the basic fact that
having low numbers of infections and deaths
does not necessarily mean
having the epidemic under control.
\nl
\mbox{} \vspace{-0.250cm} \\
%

%
%
%
%
\mbox{} \hfill 
\begin{minipage}[t]{9.10cm}  
\includegraphics[keepaspectratio=true, scale=0.70]{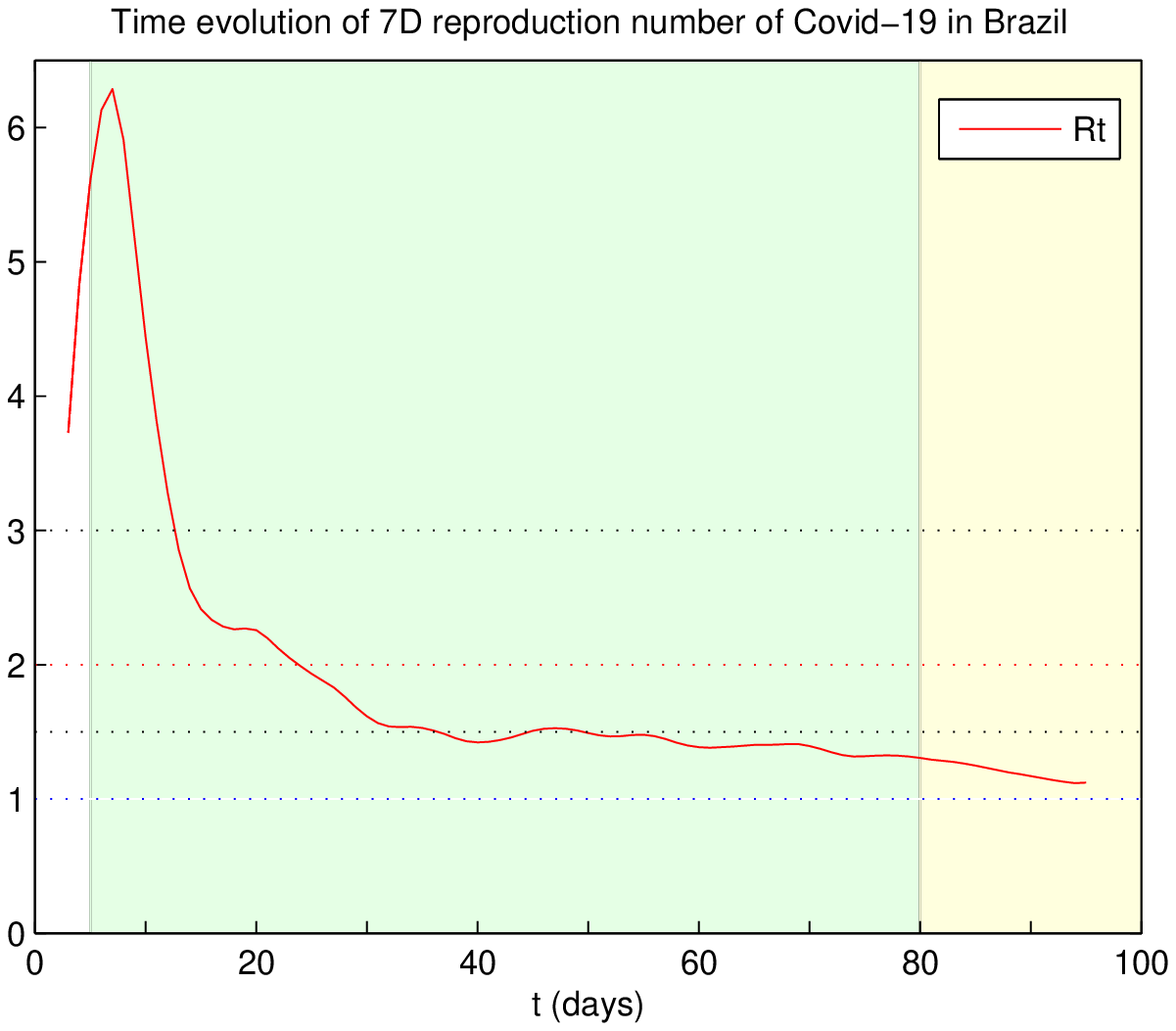}
\end{minipage}
\mbox{} \vspace{-7.850cm} \\
\mbox{} \hspace{-0.250cm}
\begin{minipage}[t]{6.000cm}
{\small \bf Example 2:}
{\small
Time evolution of Co\:\!- \linebreak
vid-19 in Brazil since 03/13/2020, \linebreak
the date of 98 total cases reported
($ t = 0 $).
With a poor coordination
between the central and
regional authorities
and different levels
of intervention in
the various states
of the country,
the decreasing of $R_t$
after reaching 1.5
by mid-April
proceeded very slowly
(green band)
due to the spread
of the epidemic
and the emergence
of new infection foci. \linebreak
\mbox{Relaxation\:measures\:began\:to\:be\:im-}
plemented
on different dates accord-  \linebreak
}
\end{minipage}
\mbox{} \vspace{-0.325cm} \\
{\small
ing to the individual regions,
but can be traced back to
06/01/2020 ($ t = 80$)
on the average.
Despite
the encouraging
behavior of $R_t$
shown in the last fortnight
(yellow band),
the indicator is likely to
resume increasing
due to further disease development
in less affected areas of the country,
particularly the southern and central
western states.
Another negative factor
is that flexibilization of control measures
has been introduced before the various regions
had attained a state of epidemic control
($ R_t \!\;\!<\!\:\! 1$),
which is {\em not\/} ideal. \\
}
\nl
%

%
%
%
%
\mbox{} \hfill 
\begin{minipage}[t]{9.10cm}  
\includegraphics[keepaspectratio=true, scale=0.70]{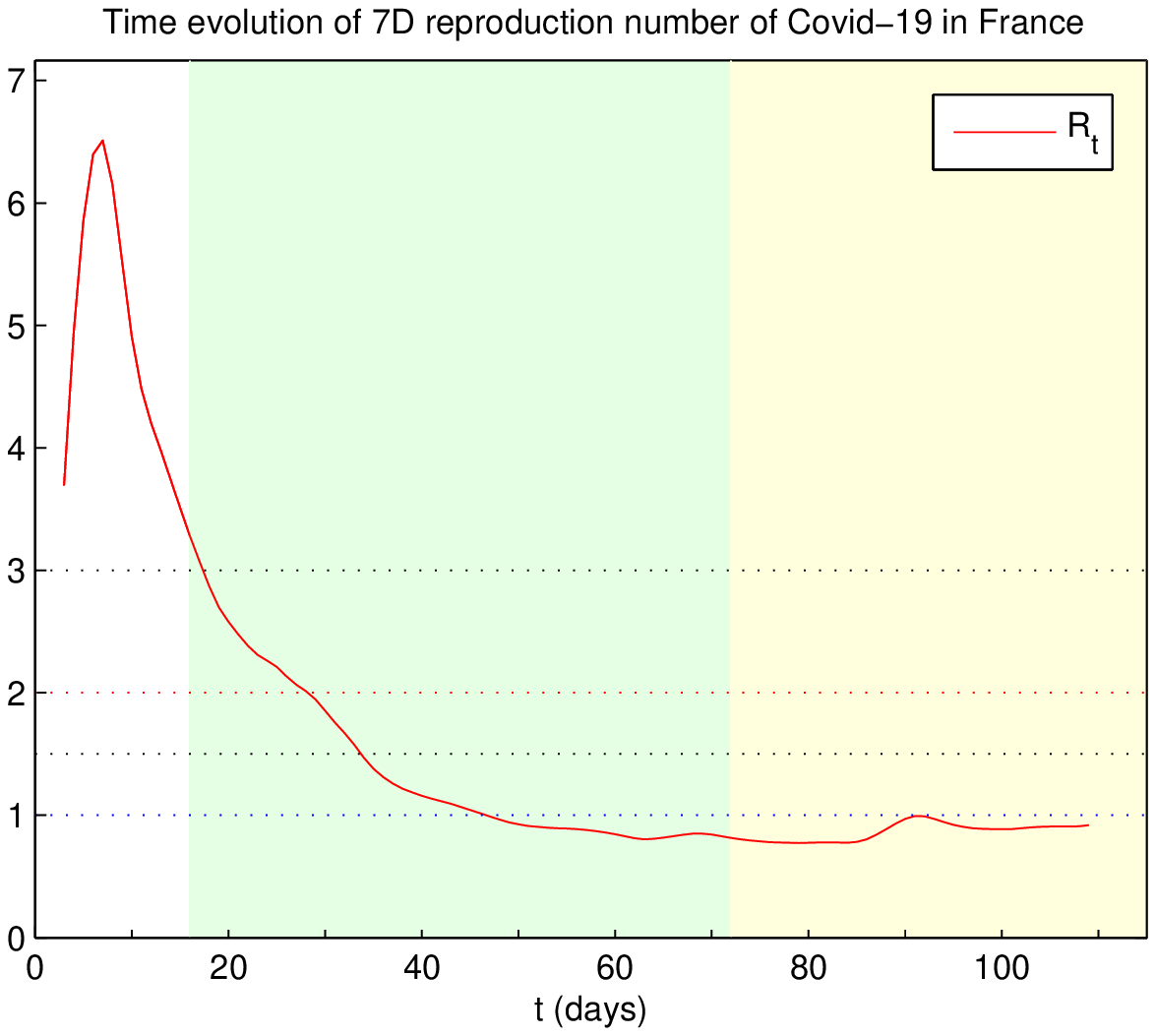}  
\end{minipage}
\mbox{} \vspace{-7.850cm} \\
\mbox{} \hspace{-0.250cm}
\begin{minipage}[t]{6.000cm}
{\small \bf Example 3:}
{\small
Time evolution of Co\:\!- \linebreak
vid-19
in France since 02/29/2020
($ t = 0 $), the date
of 100 total cases reported.
\!Containment measures
began relatively late on 03/16/2020 ($t = 16$),
with a strict eight-week lockdown
that reduced
the value of $R_t$
down to 0.81
(green band).
Restrictions were afterwards relaxed
(yellow band), with $ R_t $
stable for a couple of weeks,
when it began increasing.
A peak value of 0.99
was reached on 05/30/2020,
followed by a
reduction to its present value 0.92. \linebreak
}
\end{minipage}
\mbox{} \vspace{-0.325cm} \\
{\small
The situation requires constant monitoring,
with the possibility of having to reimpose some
restrictions to keep the epidemic under control
($ R_t \!\;\!<\!\:\!1 $). \\
}
\nl
%

%
%
%
%
\mbox{} \hfill 
\begin{minipage}[t]{9.10cm}  
\includegraphics[keepaspectratio=true, scale=0.70]{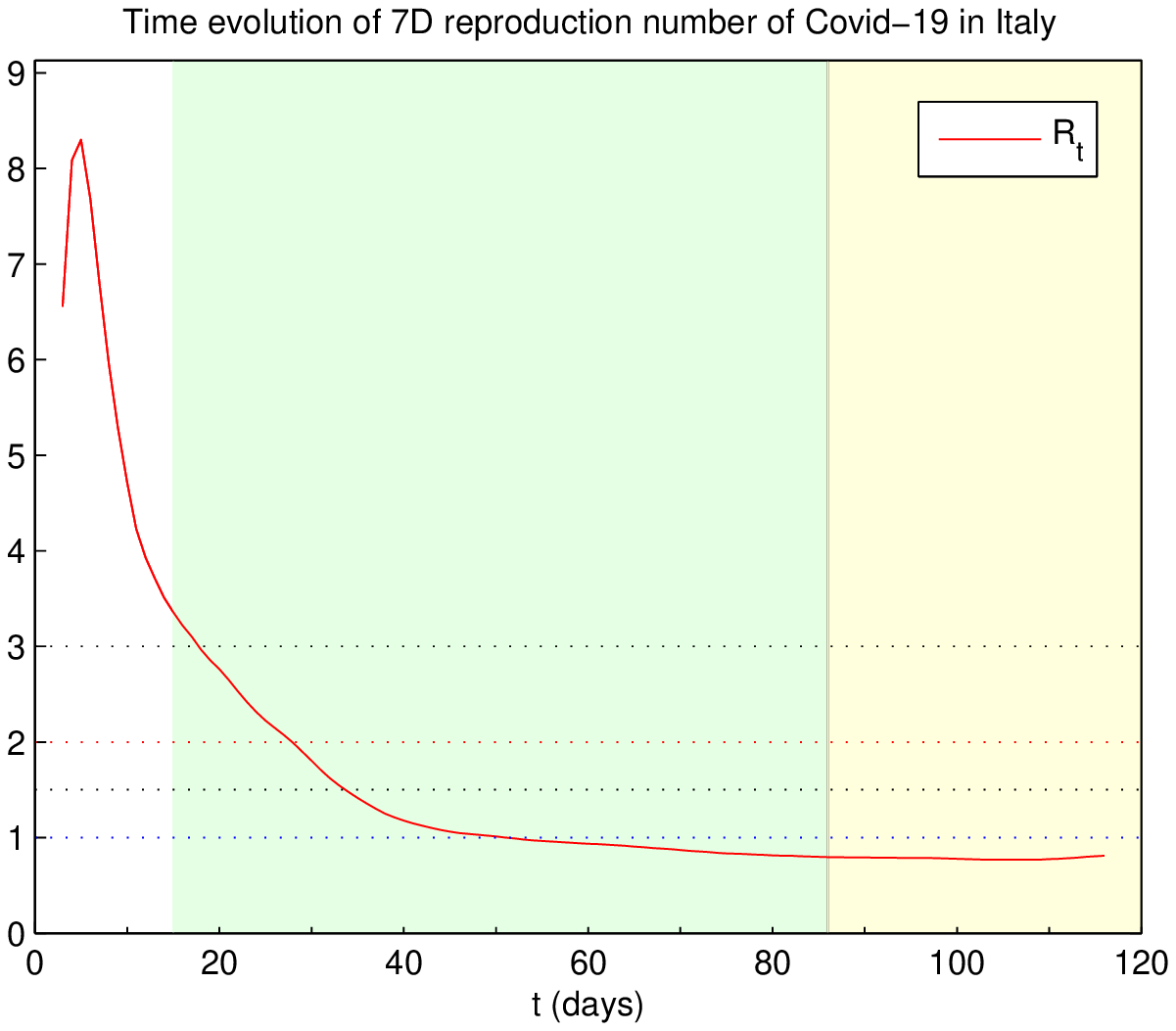}
\end{minipage}
\mbox{} \vspace{-7.850cm} \\
\mbox{} \hspace{-0.250cm}
\begin{minipage}[t]{6.000cm}
{\small \bf Example 4:}
{\small
Time evolution of Co\:\!- \linebreak
vid-19
in Italy since 02/22/2020,
the date
of 79 total cases reported
($ t = 0 $).
\!Containment measures
be\:\!- \linebreak
gan fifteen days later,
with a strict eight-week
national lockdown
imposed on 03/10/2020
($ t = 17 $).
The strong intervention
succeeded \linebreak
in continually
reducing $ R_t $
down to a safe value
of 0.80 on 05/18/2020
($ t = 86 $),
when some of the conten\:\!- \linebreak
tion rules
began being relaxed
(yel\:\!- \linebreak
low\;\!\;\!band).\;\!\mbox{The\:descent\,continued\:for} \\
nineteen days,
reaching a bottom \linebreak
}
\end{minipage}
\mbox{} \vspace{-0.325cm} \\
{\small
value of
0.77 on 06/06/2020
($ t = 105 $).
After this,
a steady and very slow increase
set in leading
to the present value of 0.81
($ t = 116 $). \\
}
\nl

%
%
%
%
\mbox{} \hfill 
\begin{minipage}[t]{9.10cm}  
\includegraphics[keepaspectratio=true, scale=0.70]{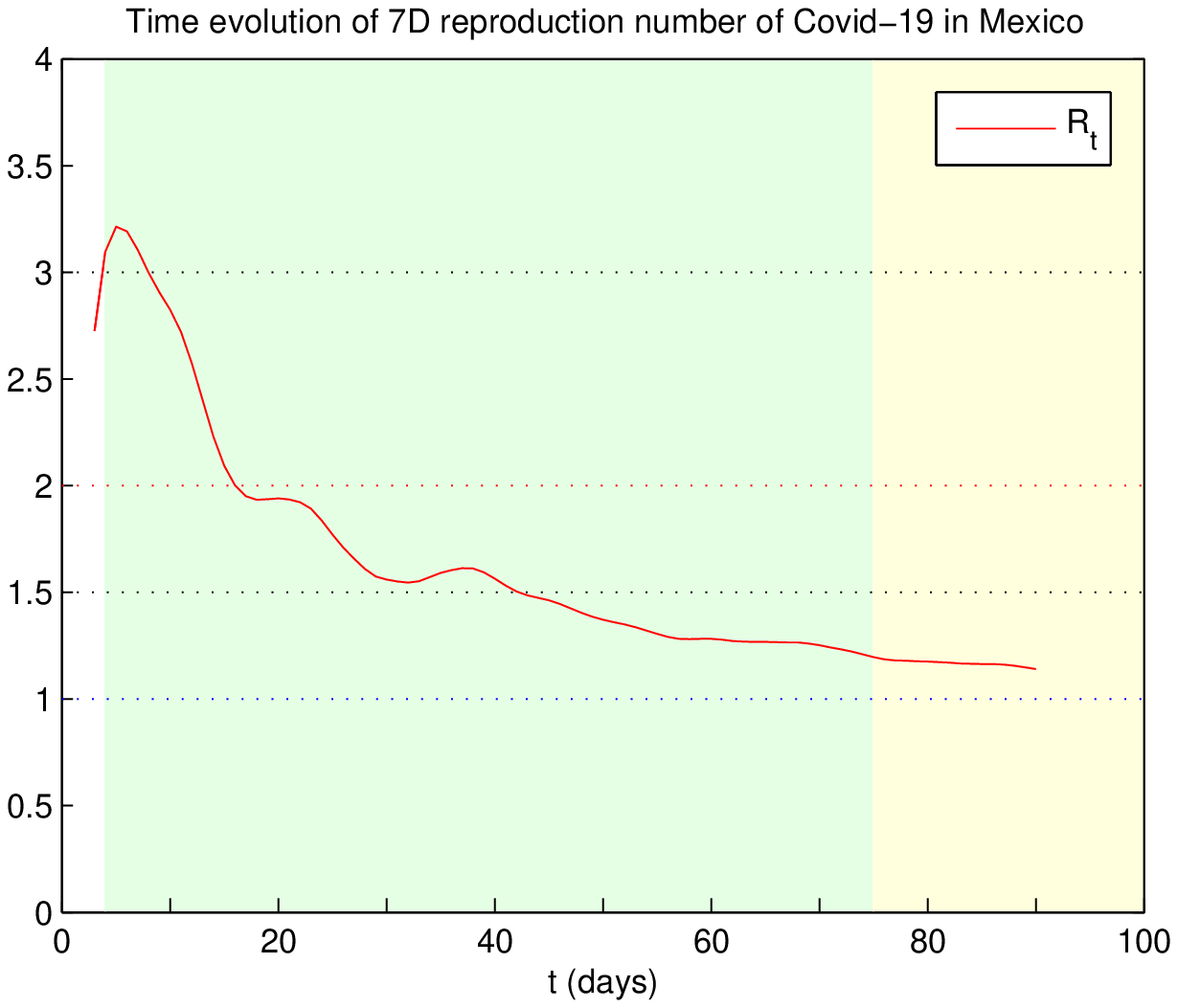}
\end{minipage}
\mbox{} \vspace{-7.850cm} \\
\mbox{} \hspace{-0.250cm}
\begin{minipage}[t]{6.000cm}
{\small \bf Example 5:}
{\small
Time evolution of Co\:\!- \linebreak
vid-19
in Mexico since 03/18/2020,
the date
of 93 total cases reported
($t = 0 $).
\!After containment measures
began on 03/22/2020 ($t = 4$),
the value of $R_t$ continually
decreased
to 1.20 (green band),
when restrictions
began to be relaxed on 06/01/2020
(yellow band).
Relaxing measures have seemingly
not changed
the behavior of $R_t$ afterwards,
but reaching a state of control
($R_t < 1 $)
still looks far away.
Similarly to Argentina and Brazil, \linebreak
}
\end{minipage}
\mbox{} \vspace{-0.350cm} \\
{\small
the flexibilization
started before the country had properly
entered the safe zone
\mbox{$ R_t \!\:\!<\!\;\! 1 $}. \\
}
\nl
\mbox{} \vspace{-0.250cm} \\
%

%
%
%
%
\mbox{} \hfill 
\begin{minipage}[t]{9.10cm}  
\includegraphics[keepaspectratio=true, scale=0.70]{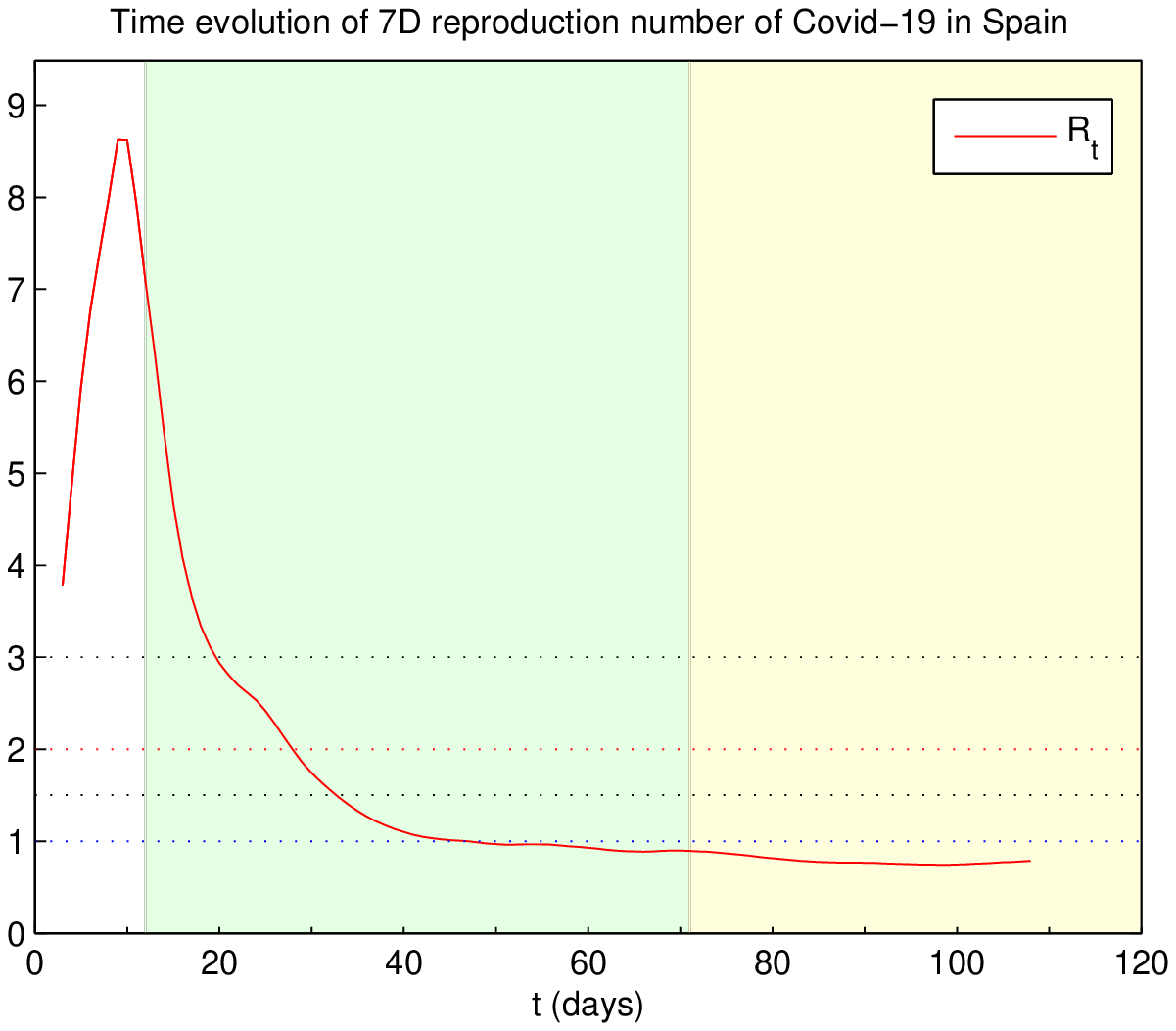}
\end{minipage}
\mbox{} \vspace{-7.850cm} \\
\mbox{} \hspace{-0.250cm}
\begin{minipage}[t]{6.000cm}
{\small \bf Example 6:}
{\small
Time evolution of Co\:\!- \linebreak
vid-19
in Spain since 03/01/2020,
the date
of 84 total cases reported
($t = 0 $).
\!After containment measures \linebreak
began on 03/13/2020 ($t = 12$),
the value of $R_t$ continually
decreased
to 0.89 on 05/11/2020 ($ t = 71 $),
when restrictions
began to be relaxed
(yellow band).
\!A minimum value of\;\! 0.74
was finally reached on 06/07/2020
($ t = 98 $),
after which a slow, steady
increase set in
towards the present value
of\;\! 0.79 ($ t = 108 $),
in a similar way to Italy. \\
}
\end{minipage}
\mbox{} \vspace{-0.350cm} \\
\nl
%

%
%
%
%
\mbox{} \hfill 
\begin{minipage}[t]{9.10cm}  
\includegraphics[keepaspectratio=true, scale=0.70]{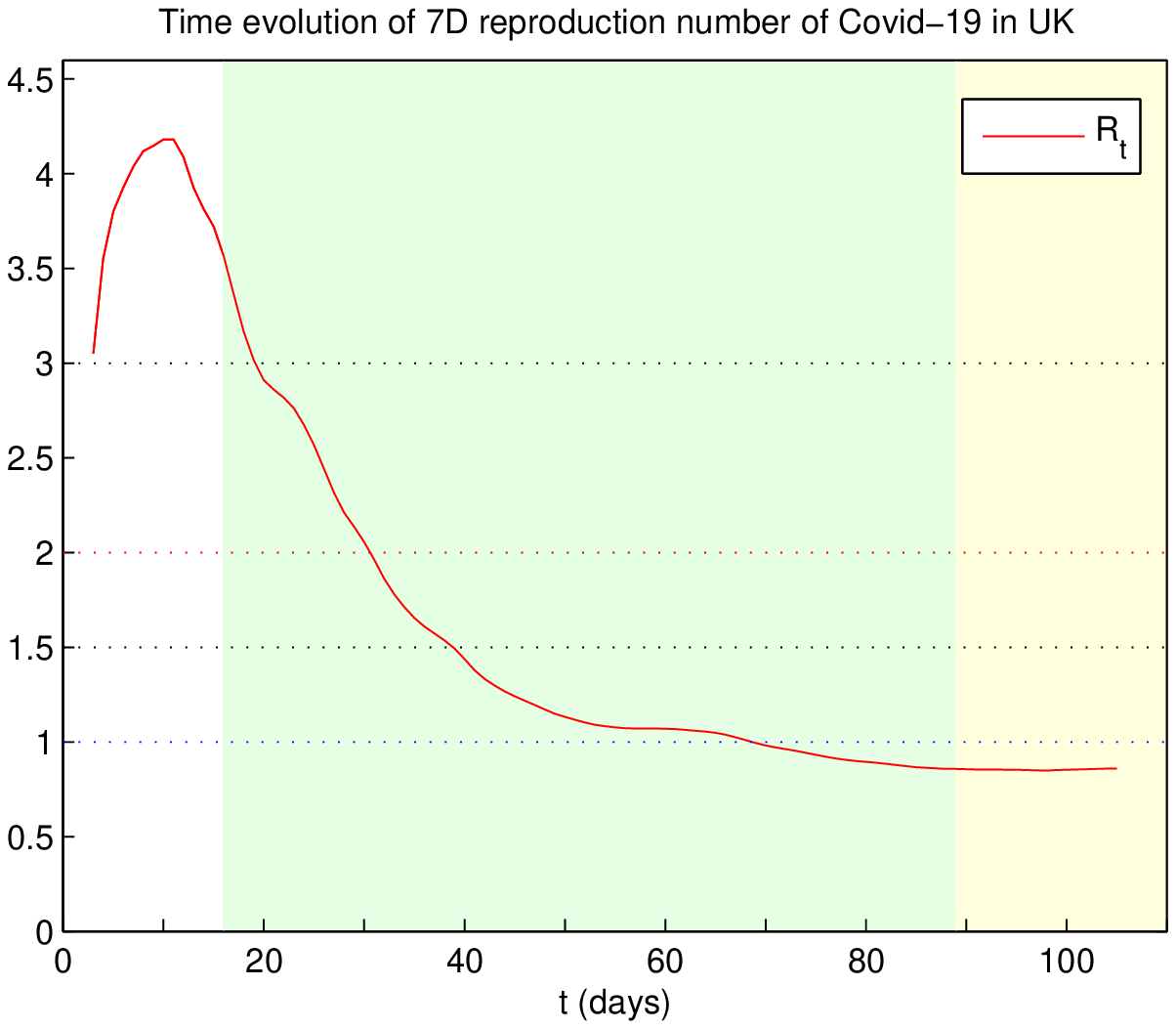}
\end{minipage}
\mbox{} \vspace{-7.850cm} \\
\mbox{} \hspace{-0.250cm}
\begin{minipage}[t]{6.000cm}
{\small \bf Example 7:}
{\small
Time evolution of Co\:\!- \linebreak
vid-19
in the UK since 03/04/2020,
the date
of 87 total cases reported
($t = 0 $).
\!After containment measures \linebreak
began relatively late
on 03/20/2020 ($t = 16$),
including strict national lockdown
and other rules
three days later,
the value of $R_t$
continually
decreased
to 0.98 on 05/13/2020 ($ t = 70 $),
when restrictions
began to be relaxed,
and then further down \linebreak
to 0.86
nineteen days later,
when the \linebreak
\mbox{lockdown\,was\,removed\;\!(yellow\,band)}.
\mbox{Despite\:successfully\,bringing\:the\:epi\:\!-} \\
}
\end{minipage}
\mbox{} \vspace{-0.350cm} \\
{\small
demic under control, the number of reported cases and deaths
was very high due to the initial delay in taking
intervention action. \\
}
\nl
%

%
%
%
%
\mbox{} \hfill 
\begin{minipage}[t]{9.10cm}  
\includegraphics[keepaspectratio=true, scale=0.70]{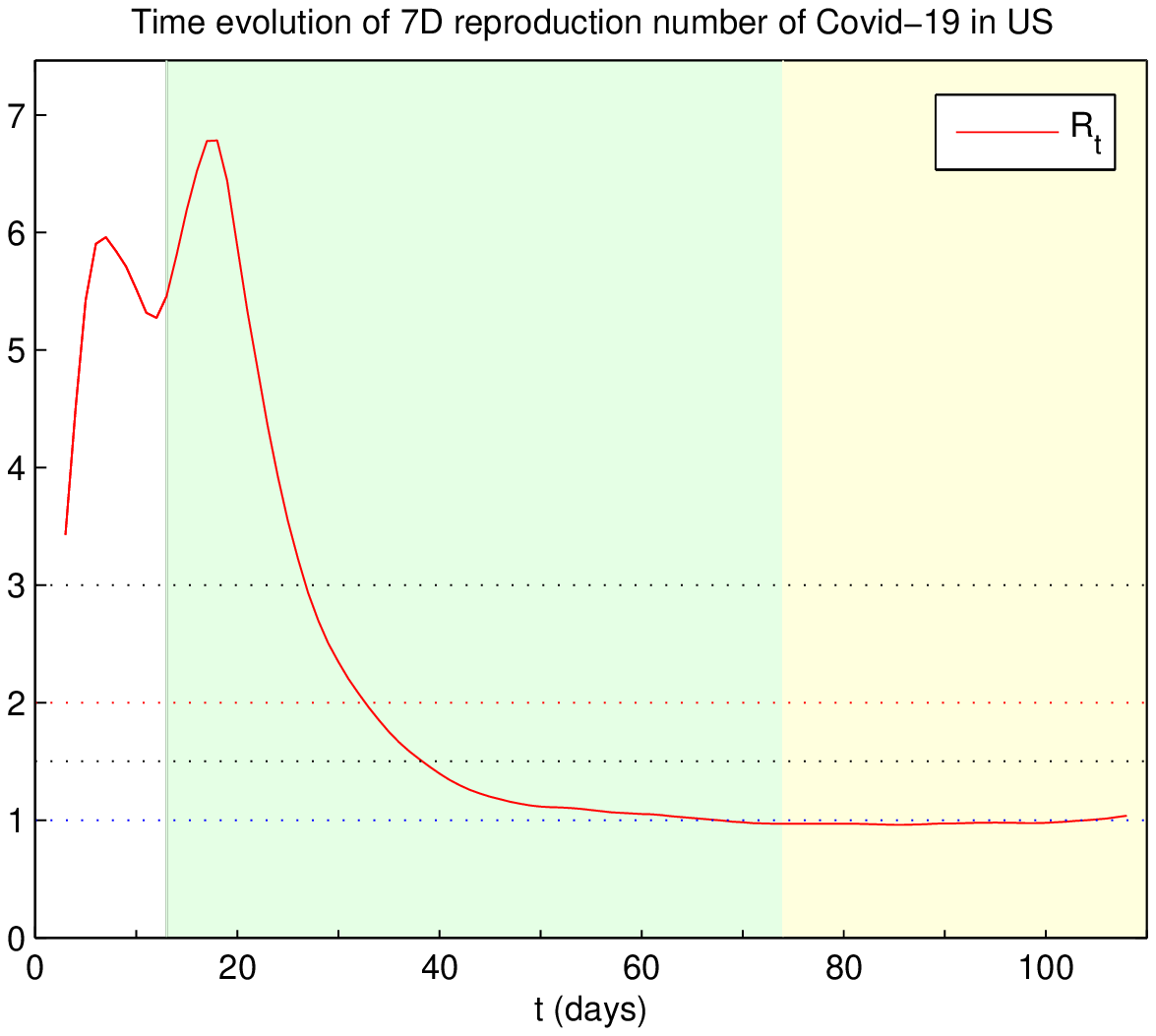}
\end{minipage}
\mbox{} \vspace{-7.850cm} \\
\mbox{} \hspace{-0.250cm}
\begin{minipage}[t]{6.000cm}
{\small \bf Example 8:}
{\small
Time evolution of Co\:\!- \linebreak
vid-19
in the US since 03/02/2020,
the date
of 100 total cases reported
($t = 0 $).
\!After containment measures \linebreak
began
on 03/15/2020 \!\mbox{($t = 13$)},
$\!R_t$ \linebreak
successfully
decreased
continually to \linebreak
0.97 on 05/15/2020 ($ t = 74 $),
when \linebreak
restrictions
began to be relaxed, \linebreak
and then slightly down
to 0.96
on 05/27/2020 ($t = 86 $),
followed by
a slow and steady ascent
to the present value of 1.04
(yellow band).
With a poor coordination between \linebreak
central and local authorities in the \linebreak
}
\end{minipage}
\mbox{} \vspace{-0.350cm} \\
{\small
beginning,
the country suffered a high mortality rate
($ 0.037\,\%$) and number of infections
(2.4 million cases reported).
Despite their efforts,
the United States have not yet
succeeded in bringing the epidemic
under nationwide control. \\
}
\mbox{} \vspace{-0.350cm} \\
\nl

\nl
\nl

%
%

%
%

\end{document}